\documentclass{jfm}

\usepackage{graphicx}
\usepackage{newtxtext}
\usepackage{newtxmath}
\usepackage{natbib}
\usepackage{hyperref}
\hypersetup{
    colorlinks = true,
    urlcolor   = blue,
    citecolor  = black,
}

\newcommand{\RomanNumeralCaps}[1]
\linenumbers

\usepackage{epstopdf, epsfig}
\usepackage{booktabs}
\usepackage{color}
\usepackage{mathtools}
\usepackage{ amssymb }
\usepackage{subcaption}
\usepackage{bm}
\usepackage{verbatim}
\usepackage[dvipsnames]{xcolor}

\newcommand{\bpu}{{\bm{\tilde{u}}}}
\newcommand{\pu}{{\tilde{u}}}
\newcommand{\pv}{{\tilde{v}}}
\newcommand{\pw}{{\tilde{w}}}
\newcommand{\pp}{{\tilde{p}}}
\newcommand{\su}{{\check{u}}}
\newcommand{\sv}{{\check{v}}}
\newcommand{\sw}{{\check{w}}}
\newcommand{\bmnu}{\bm{\tilde{u}}}
\newcommand{\mnu}{\tilde{u}}
\newcommand{\mnv}{\tilde{v}}
\newcommand{\mnw}{\tilde{w}}
\newcommand{\mnp}{\tilde{p}}

\newcommand{\bsu}{{\bm{\check{u}}}}

\newcommand{\bx}{\hat{\mbox{\boldmath$x$}}}
\newcommand{\bU}{{\bf \cal U}}
\newcommand{\bu}{{\bf u}}
\newcommand{\U}{{\cal U}}
\newcommand{\eps}{\varepsilon}
\newcommand{\bnab}{\mbox{\boldmath$\nabla$}}
\newcommand{\bLambda}{\mbox{\boldmath$\Lambda$}}
\newcommand{\bA}{{\bf C}}

\newcommand{\bF}{{\bf F}}
\newcommand{\bq}{{\bf q}}
\newcommand{\bFH}{{\bf F^H}}
\newcommand{\Rt}{Re_{\tau}}

\title{Threshold transient growth as a criterion for turbulent mean profiles}

\author{Vilda K. Markeviciute\aff{1}
  \and
  Rich  R. Kerswell\aff{1}
 }

\affiliation{\aff{1}Department of Applied Mathematics and Theoretical Physics, University of Cambridge, Wilberforce Rd, Cambridge CB3 0WA, UK}

\begin{document}

\maketitle
 

\begin{abstract}

Lozano-Duran et al ({\em J. Fluid Mech.}, {\bf 914}, A8, 2021) have recently identified the ability of streamwise-averaged turbulent streak fields $\U(y,z,t)\bx$  in minimal channels to produce short-term transient growth as the key linear mechanism needed to sustain turbulence at $Re_{\tau}=180$. 
Here, in an attempt to extend this result to larger domains and higher $\Rt$, we model this streak transient growth as a two-stage linear process by first selecting the dominant streak structure  expected to emerge over the eddy turnover time on the turbulent mean profile $U(y)\bx$, and then examining the secondary growth on this (frozen) streak field  $\U(y,z)\bx$.
Choosing the mean streak amplitude and eddy turnover time  consistent with simulations captures the growth thresholds found by Lozano-Duran et al. (2021) for sustained turbulence. 
In a larger domain at $\Rt=180$, the most energetic near-wall streaks observed in simulations are close to the predicted optimal streaks. This most energetic streak spacing, approaches the optimal streak at $\Rt=550$ where the secondary growth possible on each also comes together. 
A key prediction from the model is that the threshold transient growth required to sustain turbulence decreases with increasing $\Rt$.
More fundamentally, the work of Lozano-Duran et al. (2021) and our results suggest a subtle but significant revision of Malkus's ({\em J. Fluid Mech.}, {\bf 521}, 1, 1956) classic hypothesis concerning realisable turbulent mean profiles. The key property for  a realisable turbulent mean profile could be the ability to generate sufficient short-term transient growth rather than dependence on its (long-term) linear stability characteristics which was Malkus's original idea.  
\end{abstract}


\section{Introduction}

%
%
In turbulent wall-bounded shear flows, only the mean flow is energised by external driving effects and so it has to pass on some of its energy to the fluctuation field. Linearising the Navier-Stokes equations around the mean velocity profile is sufficient to capture all the possible energy transfer processes. Hence, linear models based on modal stability analysis or transient growth are popular tools used in flow control \citep{Kim07,Rowley17}.
Recently, in a comprehensive cause-and-effect study, \cite{Lozano-Duran21} (hereafter LD21) identified that transient growth around the streamwise-averaged velocity profile is the essential linear mechanism needed to sustain turbulence in plane channel flow with a particular threshold of the growth required at $Re_\tau=180$ in a minimal flow unit. 
This result hints at a possible and interesting update on Malkus's (1956) well-known but  defunct hypothesis \citep[e.g.][]{Tiedermann67} that the turbulent mean profile $U(y)\bx$ (where $x$ is the streamwise and $y$ the cross-shear directions) is marginally (linearly) stable. Instead, the result of LD21 suggests that there may be some sort of statistical (linear) transient growth threshold on the extended mean profiles $\U(y,z,t)\bx$ (where $z$ is the spanwise direction) realised in the flow (`statistical' here means in some averaged sense over the family of realised profiles $\U(y,z,t)$ parametrised by time $t$ rather than `statistical stability' ideas considered in \cite{MK23}). An alternative perspective is that the mean $U(y)\bx$  has a threshold for {\em nonlinear} transient growth where the growth possible for perturbations of a finite amplitude have to be considered (`finite' because the streak field is generated as part of the growth.

The work of LD21 needs extending, however, to larger flow domains to test robustness and higher $Re_\tau$ to reveal how the threshold growth possibly scales.
Unfortunately, the amount of computation involved is forbidding which suggests trying to identify and study a theoretical proxy. So motivated, we build a simple two-stage model of primary and secondary linear transient growth here and use this to suggest how LD21's results would generalise.
The primary linear process is the energy transfer from the streamwise- and spanwise-averaged velocity profile $U(y)\bx$ via streamwise rolls to spanwise-dependent but streamwise-independent streaks $\bU(y,z,t)\bx$   due to the non-normality of the linear operator. 
This transient growth process is now well understood (e.g. \cite{Orr1907, farrell_1988} for laminar flows and \cite{Kim_2000,DelAlamo06, Cossu09} in turbulent settings). 
The streak formation can also be explained through a stable mode in statistically forced turbulence modelled with Statistical State Dynamics (SSD) \citep{Farrell17}, or by the pattern forming properties of the lift-up, shear and diffusion of the mean profile \citep{Cherny05}.
Most notably, \cite{Butler93} were able to show that the observed spanwise streak spacing was consistent with optimal disturbances constrained to grow maximally over an eddy turnover time.

%
%

Taking the new base flow as the final primary streak structure added to  the mean velocity profile, the secondary linear process can be considered to model the subsequent streak breakdown. The exact linear mechanism driving the breakdown has been widely discussed.
While some studies emphasised the importance of the modal instability of the streaks \citep{hamilton_kim_waleffe_1995,andersson_brandt_bottaro_henningson_2001}, \cite{Schoppa_Hussain_2002} showed that most of the streaks observed in simulations are in fact exponentially stable and suggested transient growth as the driving mechanism of streak breakdown.
The feasibility of this conclusion was debated \citep{Jimenez2018} and alternative explanations such as parametric streak instability considered \citep{Farrell99,Farrell16}.
Recently, in their exhaustive cause-and-effect study of possible secondary linear mechanisms, LD21 showed that transient growth fuelled by `push-over' and Orr mechanisms \citep{Orr1907} is the necessary ingredient in sustaining turbulence in minimal channels at least at $Re_\tau=180$.
However, it remains unknown what perturbations and primary streak amplitudes lead to transient growth levels sufficient to sustain turbulence, the understanding of which could enhance flow control strategies. 
Previous work in this direction is mainly DNS-based including the direct-adjoint looping optimisation of the non-linear Navier-Stokes equations around a laminar Poiseuille profile with varying amplitude streak \citep{cossu_et_al_2007}.

%
%
In this paper we examine a two-stage transient growth process to study the optimal perturbation energy gain starting from a $U(y)\bx$ mean flow generated by DNS in a minimal channel. A (primary) transient growth calculation is then performed in the same spirit as \cite{Butler93} which seeks the maximal streak produced over the local eddy turnover time $T$, viz
\begin{equation}
{\rm Primary \,\,growth}: \quad U(y)\bx +\eps \bpu(y,z,0) \xrightarrow{\text{max}_{\bpu(y,z,0)} E_p}  U(y)\bx+\eps \bpu(y,z,T)
\end{equation}
where $E_p:= \int_V \bpu(y,z,T)^2 d^3 {\bf x}$ and $\int_V  \bpu(y,z,0)^2 d^3 {\bf x}=1$ and $\eps \rightarrow 0$ is assumed. This selects a unique streak structure $\bpu(y,z,T)$ but not an amplitude since only $O(\eps)$ terms are retained. We then examine the (secondary) transient growth possible on a new base flow 
of mean plus frozen, final streak field across an array of streak amplitudes $A$ - i.e.
$\U(y,z;A):= U(y) \bx+A\bpu(y,z,t_p)$ - 
as a function of $t$
(see expression given in (\ref{streaks}) below), viz
\begin{equation}
{\rm Secondary \,\,growth}: \quad \U(y,z;A)\bx +\eps \bsu(x,y,z,0) \xrightarrow{\text{max}_{\bsu(x,y,z,0)} E_s}  \U(y,z;A)\bx+\eps \bsu(x,y,z,t)
\end{equation}
where $E_s:= \int_V \bsu(x,y,z,t)^2 d^3 {\bf x}$ and $\int_V \bsu(x,y,z,0)^2 d^3 {\bf x}=1$.
This simple process is found to capture transient growth of the perturbations consistent with observations in LD21 at $Re_\tau=180$ when streak amplitudes seen in the DNS and the appropriate time horizon are used. 
We then  exploit this correspondence to predict what sort of transient growth can be achieved  in larger channels and higher Reynolds numbers.



\section{Problem set-up}
\label{sec:2}

%
%
\subsection{Mean velocity profile}

To obtain the mean velocity profile for the primary transient growth calculations, either direct numerical simulations were performed in the minimal channel flow unit \citep{jimenez_moin_1991} using the DNS code Dedalus \citep{dedalus} or data was imported from the large channel turbulent runs of \cite{lee_moser_2015} (hereafter LM15).

\begin{figure}
    \centering
    \includegraphics[width=0.7\textwidth]{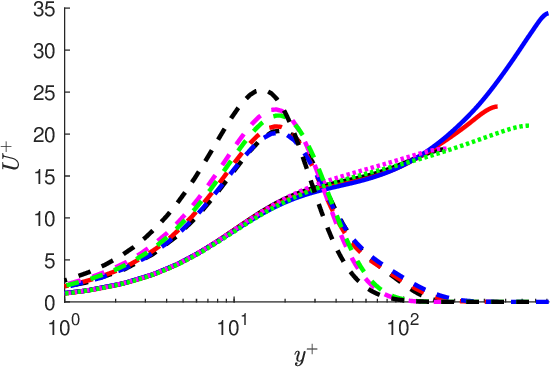}
    \caption{ 
     $U^+(y^+)$ (black, red and blue solid lines for minimal channels at $Re_\tau=180, 360$ and $720$ respectively, magenta and green dotted lines for LM15 channels at $Re_\tau=180$ and $550$ respectively). The primary streak profiles $A \mnu_{n_p}(y;t_p)$ (see expression (\ref{streaks}) below) with $n_p=1$ or $\lambda_z^{p+}=168$ for the minimal and $\lambda_z^{p+} = 154 (136)$ for the large channels at $Re_{\tau} = 180 (550)$ respectively, are shown using dashed lines and the corresponding colours with $A=10 / \sqrt{Re_{\tau}}$ to highlight their similarity ($\mnu_{n_p}(y;t_p)$ is normalised to have the same amplitude as $U$ which scales with $\sqrt{Re_\tau}$ in viscous units). The primary streaks are all peaked at $y^+ \approx 18$ consistent with the findings of \cite{Butler93}.} 
    \label{fig:profiles}
\end{figure}

The streamwise, wall-normal and spanwise directions of the channel are labelled as $x,y$ and $z$ respectively with corresponding velocity fields $u,v,w$ 
and pressure $p$. Variables are non-dimensionalised by the channel half-height $h$ and the average wall shear velocity $u_{\tau}$, which together with kinematic viscosity $\nu$ define the Reynolds number 
$$
Re_{\tau} \vcentcolon =  u_{\tau}h/\nu
$$
{\rm and a time unit $h/u_\tau$}.
The incompressible Navier-Stokes equations are then 
\begin{align}
    \frac{\partial \bu}{\partial t} + \bu \cdot \bnab \bu &= -\bnab p + \frac{1}{Re_{\tau}} \nabla^2 \bu +\bx, \\
    \bnab \cdot \bu &=0.
\end{align}
with an imposed streamwise pressure gradient $-\nabla P = \bx$. The same minimal flow domain, $(L_x^+,L_y^+,L_z^+)=(337,2Re_{\tau},168)$ (the superscript `+' indicates relative to viscous length units of $\nu/u_{\tau}$ or viscous time units of $\nu/u_\tau^2$) is used as in LD21.   No-slip boundary conditions were imposed at the channel walls located at $y=\pm 1$ and periodicity imposed in the streamwise and spanwise directions. Discretization was through a triple Fourier-Chebyshev-Fourier expansion in $x,y$ and $z$ respectively in Dedalus.
For the minimal channel, the mean velocity profile $U(y)$ was obtained by averaging over the interval $[T_0,T_0+T]$ with $T_0>0$ chosen to avoid initial transients and is shown for $Re_\tau=180,\,360$ and $720$ in Figure \ref{fig:profiles}. Also shown are the  mean profiles from the `large' channel runs of LM15 where $(L_x,L_y,L_z)=(8 \pi,2,3 \pi)$.
They agree well up to at least $y^+  \approx 50$ consistent with the estimate $0.3L_z^+$ given by \cite{Flores_Jimenez_2010} and so well above $y^+ \approx 18$   where the primary streaks are found to be located. 
The outputs of the simulations were checked by using different resolutions e.g. $N_y=90$ and $N_y = 180$ for $Re_{\tau} = 360$, $N_y=180$ and $N_y = 256$ for $Re_{\tau} = 720$ (LD21 used second order finite differences across half the channel for their calculations and $T=300$). 

%
%
\begin{table}
\begin{center}
\begin{tabular}{ccccrrccrcl}
$Re_{\tau}$ & $N_y$ &\hspace{0.75cm} & $L_x,L_y,L_z$  &  \hspace{0.75cm} $\lambda_x^+$ & \hspace{0.75cm}$\lambda_z^+$ &\hspace{0.5cm} &$t^+_e(y^+_*)$ & $\lambda_z^{p+}$ & \hspace{0.75cm} & $\left(\bar{A}\pm A_\sigma\right) \sqrt{Re_{\tau}}$ \\
\hline
$180$ &  90 &&$1.87, 2, 0.93$ & $337$ & $168$ && $114$ & $168$ && $0.9190 \pm 0.1731$\\
$360$ & 180 && $0.94, 2, 0.47$ & $337$ & $168$ && $97$ & $168$ && $0.9050\pm 0.5313$\\
$720$ & 180 && $0.47, 2, 0.23$ & $337$ & $168$ && $95$ & $168$ && $0.8399\pm 0.4481$\\\hline
$180$ & 192 && $8\pi, 2, 3\pi$ & $4,524$ & $1,696$ && $79$ & $106$ && $0.4119$  \\
$180$ & 192 && $8\pi, 2, 3\pi$ & $4,524$ & $1,696$ && $79$  & $154$ && $0.6923$ \\
$550$ & 384 && $8\pi, 2, 3\pi$ &  $13,823$ & $5,184$ && $89$   & $112$ && $0.1121$\\ 
$550$ & 384 && $8\pi, 2, 3\pi$ &  $13,823$ & $5,184$ && $89$  & $136$ && $0.1522$\\ \hline
\end{tabular}
\end{center}
\caption{
Parameters for the different cases considered. Top, results for the minimal box simulations and  bottom, results using data from \cite{lee_moser_2015}. $N_y$ is the wall-normal Chebyshev resolution used in transient growth calculations, $\lambda_x^+$ and $\lambda_z^+$ are the largest streamwise and spanwise wavelengths present, $\lambda_z^{p+}$ is the primary streak spacing and $t^+_e$ is the estimated eddy-turnover time. $y^+_*$ is the unique distance from the wall at a given $Re_\tau$ where the optimal streak has a maximal growth time equal to the local eddy turnover time. }
\label{tb:LM15}
\end{table}
%
%
%
%
%
%
\begin{figure}
    \centering
    \includegraphics[width=0.7\textwidth]{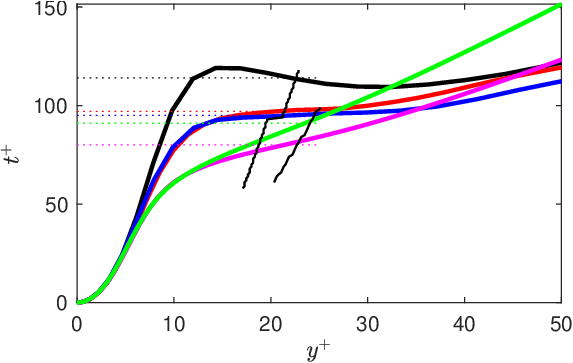}
    \caption{
    Eddy turnover time $t^+_e=Re_{\tau} q^2/\epsilon$
    (black, red and blue lines for $Re_\tau=180, 360$ and $720$ respectively, magenta and green lines for $Re^{LM15}_\tau=180$ and $550$ respectively). The location of the optimal streak as a function of the optimisation time is shown as thin black curves for the minimal channel (left) and the large channel (right) (the kink in the curve is due to the discrete change in optimal spanwise wavenumber of the streak). Chosen primary transient growth time horizons are marked with dotted lines. See Table 1 for numerical values of $t_e^+$ at the various $Re_\tau$. 
    }
    \label{fig:eddy}
\end{figure}

%
%
\subsection{Primary transient growth}

We consider primary transient growth of streamwise-independent perturbations to the mean velocity profile $U(y)$ defined as $[\pu,\pv,\pw] = [u-U,v,w]$ and decomposed into spanwise Fourier modes as follows:
\begin{equation}
        [\bpu(y,z,t), \pp(y,z,t) ] 
        = \sum_{n} [\,\bmnu_{n}(y,t), \mnp_{n}(y,t)\, ] \exp (in \beta z) + c.c.
\end{equation}
where $\beta:=2\pi/L_z$.
The linearised Navier-Stokes equations around the mean $U(y)\bx$ for each $n$ are 
\begin{equation}
\partial_t \bmnu_{n} + (\partial_y U) \mnv_{n} \bm{\hat{x}}
               = - \bm{\nabla} \mnp_{n} + \frac{1}{Re_{\tau}} \nabla^2 \bmnu_{n} , \quad \bm{\nabla} \cdot \bmnu_{n} = 0
    \label{eq:PTG}
\end{equation}
and can be treated separately.
Due to the non-normality of the linear operator, the perturbations can experience transient growth in energy quantified by the (primary) gain
\begin{equation}
    G_p(n,t_p) := \sup_{\bmnu_{n}(y,0)}\frac{||\bmnu_{n}(y,t_p)||^2}{||\bmnu_n(y,0)||^2}
\label{eq:Gp}
\end{equation}
where the energy norm is calculated by
\begin{equation}
    E_n \vcentcolon = ||\bmnu_{n}||^2 = \int_{-1}^{1}  |\mnu_{n}|^2+|\mnv_{n}|^2+|\mnw_{n}|^2 \, dy. 
    \label{energy_norm}
\end{equation}

The goal of the primary transient growth calculation is to identify the streamwise-independent 
streak field $\mnu_{n_p}(y;t_p) \exp(i n_p \beta z)$ (with wavelength $\lambda_{z}^{p}:= 2\pi/n_p \beta$) which achieves optimal energy growth at time $t_p$. This will then be used as the 2-dimensional streak field upon which to study secondary growth. Two issues require discussion: a) how to choose $t_p$ (which then specifies the streak spacing) and b) what amplitude to make the streaks for the secondary growth analysis.

%
%
\subsection{Timescale for primary transient growth} 
\label{sec:2.3}

Choosing the primary transient growth timescale $t_p$ presents a challenge as the global (over all $t_p$) optimal growth is achieved at a time much greater than the characteristic timescale of turbulent fluctuations by perturbations with larger than observed spanwise spacing.
\cite{Butler93} realised  that perturbations can only grow over the time limited by the eddy turnover time $t^+_e(y^+)=\Rt q^2(y^+)/\epsilon(y^+)$ (the ratio of the characteristic turbulent velocity squared $q^2$ and dissipation rate $\epsilon$ at a given distance $y^+$ from the wall, given in viscous units) before being disrupted by turbulent fluctuations. 
\cite{Butler93} suggested choosing $t^+_p=t^+_e(y^{+}_*)$ where $y^+_*$ is the unique wall distance where the optimal streak is positioned given a growth time equal to the local eddy turnover time.
While such restriction of the primary transient growth timescale was debated by \cite{Waleffe97} and \cite{Cherny05}, it provides a practical approach to obtain realistic streamwise streak profiles and will be used below. 
To calculate the eddy turnover time we use the following definitions:
\begin{gather}
 q^2(y):= \frac{1}{T L_z L_x } \int_{T_0}^{T_0+T} \int_0^{L_z} \int_0^{L_x} \left[ (u-U)^2 + v^2 + w^2  \right] \, dx \, dz \, dt\\
\epsilon(y) := \frac{1}{T L_z L_x } \int_{T_0}^{T_0+T} \int_0^{L_z} \int_0^{L_x} |\bnab(\bu-U\bx)|^2\, dx \, dz \, dt.
\end{gather}
Then, we compare the centre of the streak location (defined by maximum velocity) as a function of the streak optimisation time to the eddy turnover time $t^+_e$ as a function of the distance from the wall $y^+$, plotted using viscous units in Figure \ref{fig:eddy}. The intersection points are at $t^+_p=114, \,97$ and $95$ for $Re_\tau=180,\,360$ and $720$ respectively in the minimal channel (giving a streak spacing of $\lambda_z^{p+}=168$) and at $t^+_p=79$ and $t^+_p=89$ in the large channel for $Re_\tau=180$ and $550$ respectively (giving streak spacing of $\lambda_z^{p+}=106$ and $\lambda_z^{p+}=112$ respectively): see Table \ref{tb:LM15}. 

\subsection{Amplitude for the primary streak} 
\label{sec:2.4}

For the secondary transient growth calculation, the streamwise-independent base flow is defined as:
\begin{equation}
   \U(y,z) \bx :=  U(y)\bx +  \biggl[A \pu_{n_p}(y;t_p) \exp (2 \pi i z/\lambda_z^p ) +c.c \biggr]\bx
   \label{streaks}
\end{equation}
with normalisation $||\mnu_{n_p}(y;t_p)||^2=||U||^2$ (the $\pv_{n_p}$ component is not used). 
The (non-dimensional) primary streak amplitude $A$ is then the amplitude of the Fourier mode corresponding to the streak normalised by that corresponding to the mean flow.  This ratio was deduced directly from time-averaging DNS data in  the minimal channel or via the  DNS data reported by LM15 for the larger channel. Without loss of generality, the streak field is chosen to be symmetric about $z=0$ and then $\pu_{n_p}(y;t_p)$ is real. 
In the below, the mean amplitude, $\bar{A}$, and standard deviation $A_\sigma$ were collected for the minimal channel  whereas only $\bar{A}$ was available for the large channel (see Table \ref{tb:LM15}).

%
%
\subsection{Secondary transient growth}
\label{sec:stg}

Secondary transient growth perturbations are superimposed on the streak field $\U$ given in (\ref{streaks}) so that $\bsu=[\su,\sv,\sw] = [u-\U,v,w]$ where $\bsu$ is assumed infinitesimally small. The secondary perturbation is decomposed into Fourier modes,
\begin{equation}
        [\bsu(x,y,z,t), \check{p}(x,y,z,t) ] = \sum_{m} [\bsu^{m}(y,z,t), \check{p}^{m}(y,z,t) ] \exp{(i m \alpha x)} 
\end{equation}
where 
\begin{equation}
 [\bsu^{m}(y,z,t), \check{p}^{m}(y,z,t) ] = \sum^N_{n = -N} [\bsu^{m}_{n}(y,t), \check{p}^{m}_{n}(y,t) ] 
\exp{(2 \pi i(\mu+n) z/\lambda_z^{p} )},
\label{modulation}
\end{equation}
so $\mu \in [0,\tfrac{1}{2}]$ is a `modulation' parameter and $\alpha:=2\pi/L_x$. The secondary disturbance only has the same spanwise wavelength as the streak field when $\mu=0$ (no modulation) otherwise the spanwise wavelength of the secondary perturbation increases by a factor $1/\mu$ over that of the primary perturbation. 
Hereafter we just consider $\mu=0$ so the growths found below are a lower bound on what is possible across all $\mu$.

In the linearised equations determining how $\bsu$ evolves, the Fourier modes in $x$ can be treated separately but the spanwise wavenumbers are coupled by the streak field adding new terms to the linearised equations (\ref{eq:PTG}):
\begin{gather}
\begin{split}
       \partial_t \bsu^{m}_{n} &+  \overbrace{i m \alpha [U \bsu^{m}_{n} 
                              +  A \pu_{n_p}   \bsu^{m}_{n-1} 
                              +  A \pu_{n_p}^{*} \bsu^{m}_{n+1}]}^{{\rm advection}}
                +\overbrace{(\partial_y U) \sv^{m}_{n}\bm{\hat{x}}}^{{\rm lift-up}} \\
                +&\underbrace{A\left[ \partial_y \pu_{n_p} \sv^{m}_{n-1} + \partial_y \pu_{n_p}^{*} \sv^{m}_{n+1} \right] \bm{\hat{x}}}_{{\rm extra \,\, lift-up}}
                 +\underbrace{ \tfrac{2\pi i}{\lambda_z^p} A \left[\pu_{n_p} \sw^{m}_{n-1} -\pu_{n_p}^{*} \sw^{m}_{n+1}\right]\bm{\hat{x}}}_{{\rm push-over}} \\
 & \hspace{1cm}=-\left( \begin{array}{c}
i m \alpha  \\
\partial_y \\
2 \pi i n/\lambda_z^p 
\end{array}\right) \check{p}^{m}_{n}
+\frac{1}{Re_{\tau}}\biggl[\partial_y^2-m^2 \alpha^2 -\frac{4 \pi^2 n^2}{(\lambda_z^{p})^2} 
\biggr]\bsu^{m}_{n}       \label{eq:STG_1}     \end{split}\\
\nonumber \\
im\alpha \su^{m}_{n}+\partial_y \sv^{m}_{n} +\frac{2 \pi i n }{\lambda_z^p} \sw^{m}_{n}=0
\label{eq:STG_2}                  
\end{gather}
We solve equations (\ref{eq:STG_1})-(\ref{eq:STG_2}) for {\it each} value of $m$ (or equivalently wavelength $\lambda_x=2\pi/m \alpha=L_x/m$) separately while $2N_z+1$ spanwise modes are coupled using $n \in [-N_z,N_z]$. 
The definition of the secondary gain is
\begin{equation}
    G^m_s(t_s;\lambda_z^{p+},A,\Rt) := \sup_{\bsu^{m}(y,0)}\frac{||\bsu^{m}(y,z,t_s)||^2}{||\bsu^{m}(y,z,0)||^2}
\label{eq:Gs}
\end{equation}
and
\begin{equation}
G_s(t_s;\lambda_z^{p+},A,\Rt):= \max_{m}G^m_s(t_s;\lambda_z^{p+},A,\Rt).
\end{equation}
is the optimal gain across the set of wavenumbers considered (and $\mu=0$).
We explore how this optimal gain changes as the primary streak amplitude $A$ is increased over the range of values observed in the simulations. 

%
%
%
 \begin{figure}
    \centering
    \includegraphics[width=0.4\textwidth]{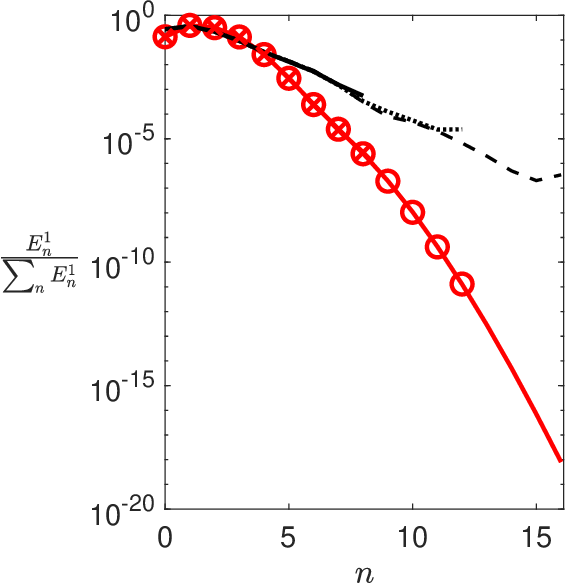}
    \caption{Energy distribution over spanwise wavenumbers $n$ for the optimal secondary transient growth mode at $\lambda_x=L_x$  and  $Re_\tau=180$ in the minimal channel with primary streak  $\lambda_z^{p+}=168$ and amplitude $\bar{A}$. Shown for $N_z = 8,12$ and $16$ using black solid, dotted and dashed lines for the input modes (respectively), and red crosses, circles and a solid line (respectively) for the output modes. This figure indicates that even for $N_z=8$, there is a spectral drop off in energy of six orders of magnitude for the output modes.}
    \label{fig:primary_res}
\end{figure}

%
%
\subsection{Numerical implementation}

To perform the primary and secondary transient growth calculations, we follow the matrix-algebra approach described in \cite{reddy_henningson_1993}. If $\bq_j=(u_j,v_j,w_j,p_j)$ is the $j$th eigenfunction corresponding to the eigenvalue $\lambda_j$ and ordered by increasing damping rate $-\Re e(\lambda_j)$, then \cite{reddy_henningson_1993} define the eigenvalue `overlap' matrix as 
\begin{equation}
C_{jk} := \langle \bq_j, \bq_k \rangle:= \int \int_{-1}^{1} \int u_j^*u_k+v_j^*v_k+w_j^* w_k\, \, dx dy dz
\end{equation}
where the inner product is that induced by the energy norm defined in (\ref{energy_norm}), the integration is over a wavelength in $x$ and $z$, and $^*$ indicates complex conjugation. This matrix is diagonal if the eigenfunctions form an orthonormal set but, as is typical for shear flow problems, is dense here. It is, however, Hermitian and positive definite which means there exists a matrix $\bF$
such that $\bA=\bFH \bF$ where $\bFH:=(\bF^*)^T$ is its Hermitian conjugate ($\bF$ is found using the SVD of $\bA={\bf U}{\bm{\Sigma}} {\bf V^H}$ so that $\bF:= {\bf U} \bm{\sqrt{\Sigma}}{\bf  V^H}$ where
$\bm{\sqrt{\Sigma}}$ indicates a diagonal matrix with the square root of $\bA$'s singular values on the diagonal.)  Then the largest growth can be computed by squaring the largest singular value of $\bF e^{t \bLambda} \bF^{-1}$ where $e^{t\bLambda}$ is the diagonal matrix exponential with 
$e^{t\lambda_j}$ on the diagonal (see equation (30) in \cite{reddy_henningson_1993}).

The symmetry in the wall-normal direction allows disturbances satisfying the symmetries 
\begin{align}
Z_{-}:(u,v,w,p)(x,y,z,t) & \to (u,-v,w,p)(x,-y,z,t) \nonumber\\
\& \quad Z_+:(u,v,w,p)(x,y,z,t) & \to (-u,v,-w,-p)(x,-y,z,t)
\end{align}
to be considered separately. 
Discretisation of  the linear operators corresponding to equations (\ref{eq:PTG}) and (\ref{eq:STG_1})-(\ref{eq:STG_2}) produce matrices of size $2N_y$ and $2N_y(2N_z+1) $ respectively. For the primary transient growth calculation, $N_e=80$ eigenfunctions were used in the expansion which was tested by accurately reproducing the results in \cite{Butler93} who used a modelled mean profile. 
Optimal streaks located near the wall were found to be the same for $Z_-$ and $Z_+$, and so the primary streak profile was chosen with $Z_-$ symmetry. No change in the optimal streaks were observed repeating the calculation with double the wall-normal resolution.

For the secondary transient growth calculation, a spanwise truncation of $N_z = 8$ proved adequate typically giving a spectral drop off of over six orders of magnitude in the output mode: see figure \ref{fig:primary_res} which compares $N_z=8$, $12$ and $16$. 
The choice of appropriate $N_e$ is more delicate with increasing values needed  for smaller $\lambda_x$ and short times. Typically $N_e=500-800$ was used as a compromise between accuracy and runtimes. In all cases, secondary  perturbations with $Z_-$ symmetry yielded dominant transient growth values and are presented in this paper. 



%
%
\begin{figure}
    \centering  
    \includegraphics[width=0.48\textwidth]{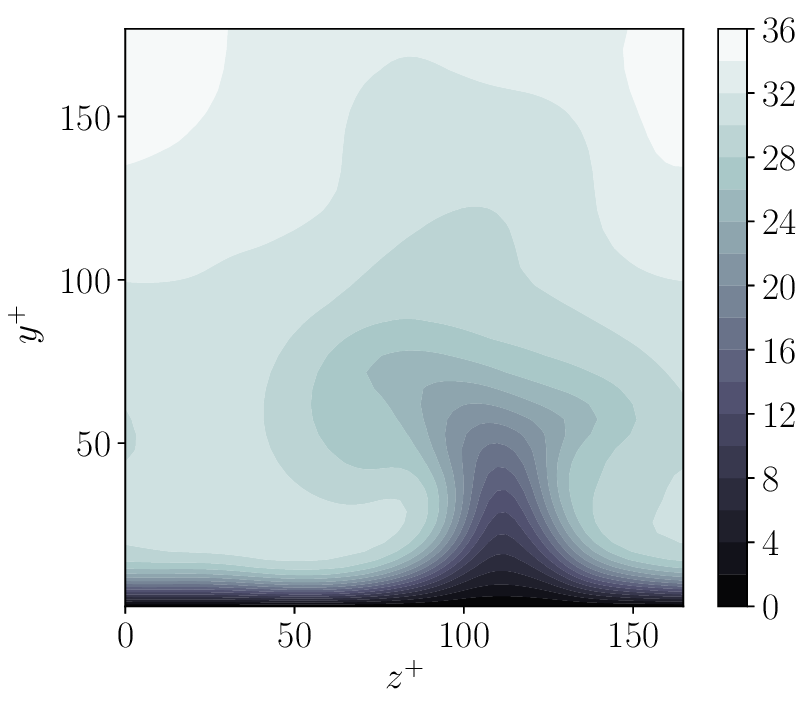}
    \includegraphics[width=0.48\textwidth]{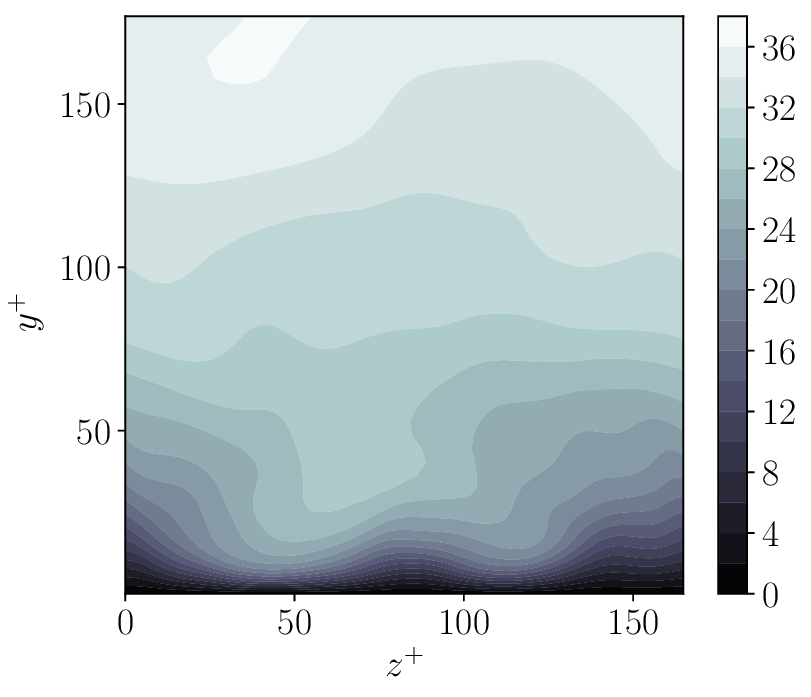}
    \caption{Sample snapshots of the $x$-averaged velocity field $U^+(y^+,z^+)$ from a $Re_{\tau}=180$ simulation in the minimal channel of LD21. Left: streaks with $\lambda^{p+}_z=168$ ($n_p=1$) and Right: streaks with 
    $\lambda^{p+}_z=84$ ($n_p=2$). These figures show that both are visible at different times in the simulations.}
    \label{fig:dns}
\end{figure}

%
%
 \begin{figure}
    \centering
    \includegraphics[width=0.47\textwidth]{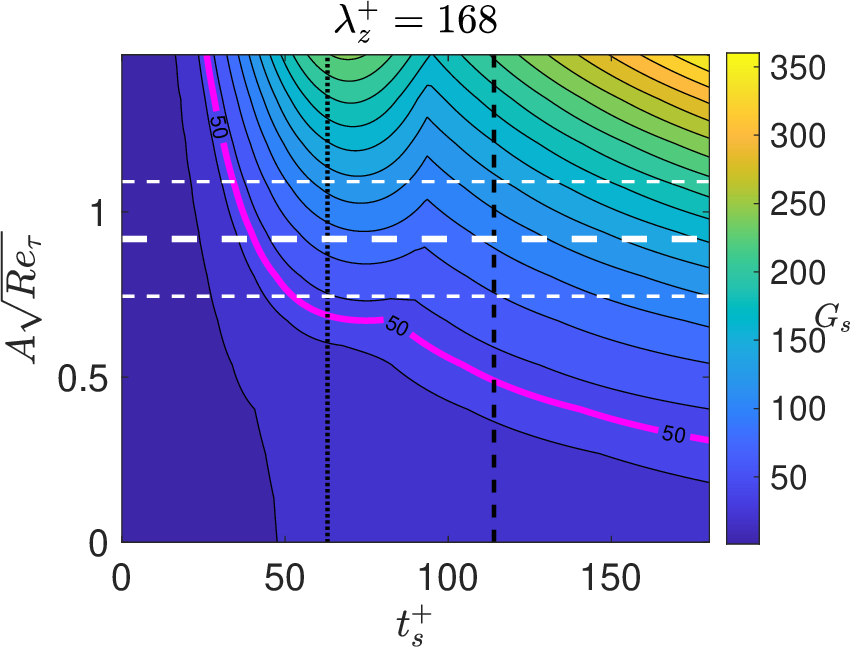}
    \includegraphics[width=0.47\textwidth]{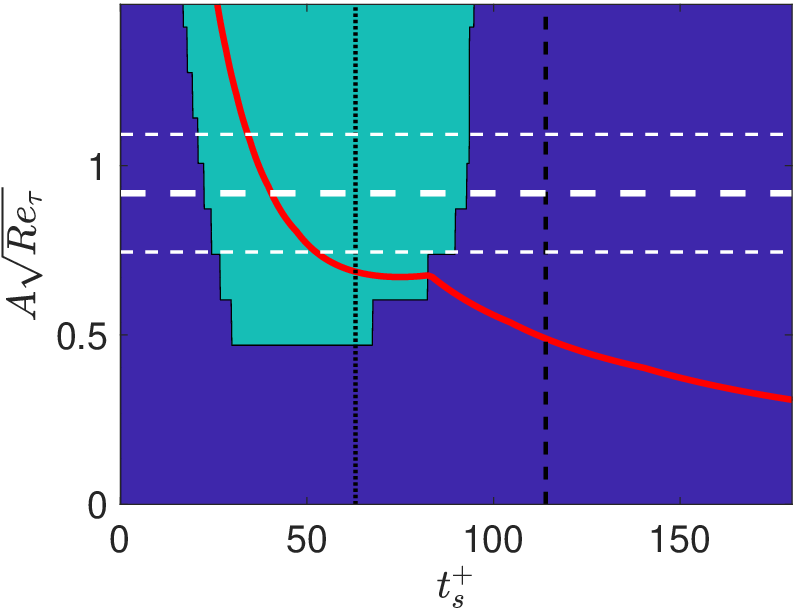}
    \includegraphics[width=0.47\textwidth]{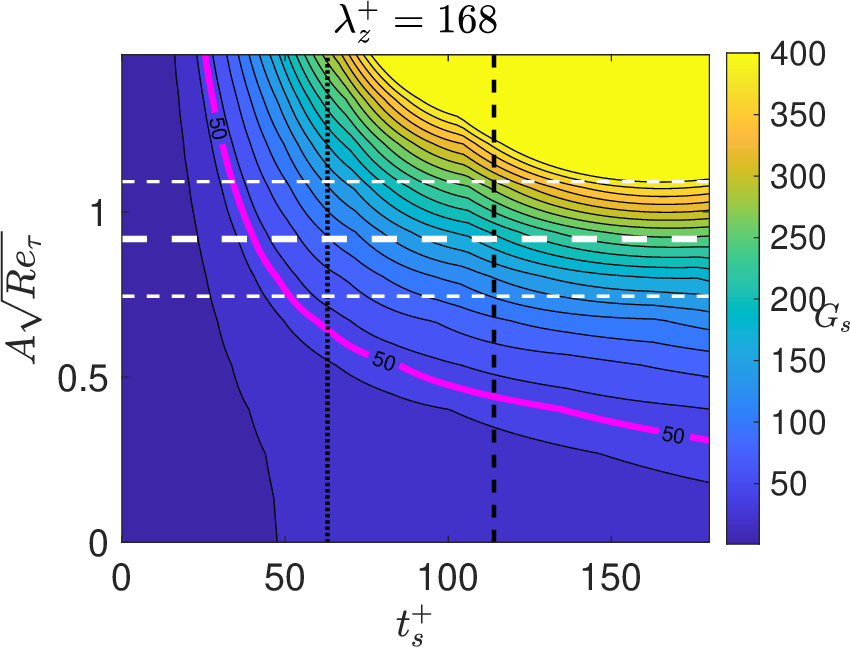}
    \includegraphics[width=0.47\textwidth]{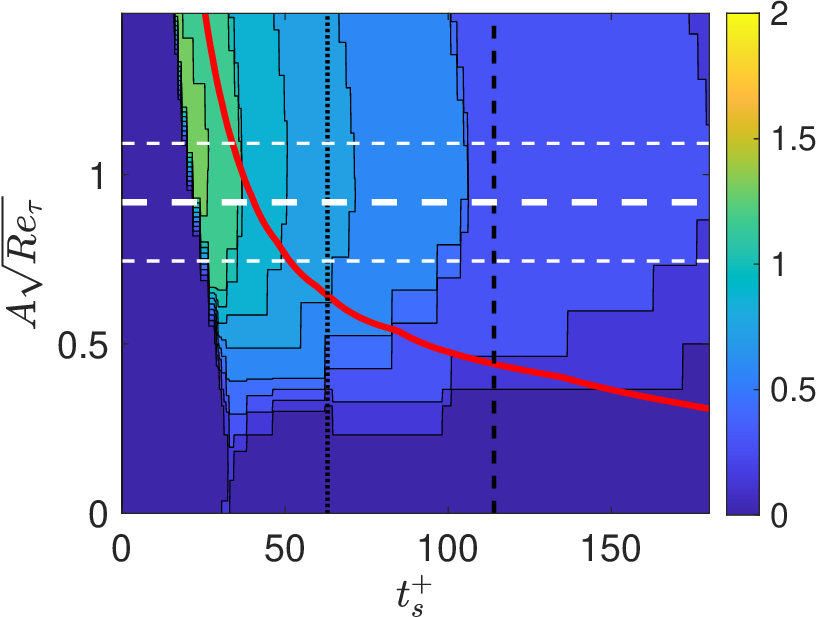}
    \caption{Top left: secondary transient growth $G_s(t_s^+;168,A,\Rt=180)$ contour plots over  the $(t_s^+,A\sqrt{Re_{\tau}})$ plane. Contour increments are 20.
    Top right: corresponding optimal streamwise wavenumber $\alpha m_{max} = 2 \pi/\lambda_x^+$, where green (blue) area corresponds to $m=1$ ($m=0$) wavenumber, showing the existence of a region where 3D perturbations become optimal. 
    The bottom left and bottom right plots show the same quantities, but 
    calculated for the set $m \in \{0,0.2,\ldots,2\}$.    
    In all four plots, the white long dashed line shows $\bar{A}$ for the streaks with standard deviations are indicated by short dashed lines (both from DNS). Time horizons $t_{LD21}^+=63$ and $t_e^+$ are marked with vertical black dotted and dashed lines respectively, and a  magenta (left) or red (right) solid contour show the $G_s((\lambda_x^+)_{\max})=50$ threshold.
    }
    \label{fig:sg_180_minimal}
\end{figure}

\section{Results}
\label{sec:3}

\subsection{\label{S3.1}Two-Stage Optimisation} 

We first consider the minimal domain and $\Rt=180$ used by LD21. The approach taken here is to do two optimisations in sequence to estimate the energy growth possible in the flow over timescales consistent with the eddy turnover times near the wall. The first optimisation is used to define the streak field using  linear transient growth analysis and the second optimisation then finds the optimal secondary growth on the primary streak field once it has been given an amplitude. This is admittedly a simplistic approach which relies on a timescale separation argument that the secondary growth occurs over a shorter timescale than the evolution of the streaks so they can be considered steady. This is unlikely to be true but the alternatives (discussed later) require a significantly more complex approach. Given this, trying this simple and interpretable approach to see if it captures the sort of growths found by LD21 seemed reasonable.

The first optimisation has already been discussed with the result that a primary streak of spanwise spacing $\lambda_z^{p+}=168$ - the largest spanwise streak which fits into the minimal box - emerges using the approach advocated by \cite{Butler93}. Reassuringly, this is the most energetic streak seen in our simulations consistent with LD21 (see their figures 2 and 3) followed by the $\lambda_z^{+}=84$ double streak  (both can be seen in the simulations: see figure \ref{fig:dns}). The results of the secondary optimal gain $G_s(t_s^+;168,A,\Rt)$ contoured over  the $(t_s^+, A \sqrt{\Rt})$ plane is shown in figure \ref{fig:sg_180_minimal} with the associated optimising streamwise wavenumber index $m$ shown on the right 
($A$ is multiplied by $\sqrt{\Rt}$ to bring it into alignment with $U$ which scales with this factor in viscous units). 
The top subfigures show the results of restricting the optimisation to integer $m$, that is,  secondary disturbances which fit into the minimal box. This shows two distinct regions where the secondary growth optimal is 2D ($m=0$) and where the optimal is 3D (always corresponding to $m=1$ so the  streak just streamwise-fits into the box).
Interestingly, for amplitudes of the primary streak seen $A \in [\bar{A}-A_\sigma,\bar{A}+A_\sigma]$ and times around that quoted by LD21, $t_s^+=t_{LD21}^+:=63$ ($=0.35h/u_\tau$ in  their figure 9), the secondary mode is 3D. In their \S6.3, LD21 also discuss a threshold of $G_s=50$ for sustainable turbulence (although their figure 22 actually suggests the threshold is closer to 40 than 50). This contour is highlighted in figure \ref{fig:sg_180_minimal}
and is clearly shown to thread the region of interest. Secondary growths exceeding this threshold  dominate this region. In particular, 
\begin{equation}
G_s(t_{LD21}^+;168,\bar{A},180) \approx 90
\end{equation}
with the growth ranging from 60 at $A=\bar{A}-A_\sigma$ to 130 for $A=\bar{A}+A_\sigma$.

The lower subfigures in figure \ref{fig:sg_180_minimal} extend the optimisation over longer streamwise wavenumbers (smaller $m$) to clarify the 2D-3D transition: the optimal disturbances become progressively longer in streamwise wavelength as the optimisation time increases whereas at short times there is a sudden transition from infinite streamwise wavelength (2D) disturbances to those of short wavelength. 

\begin{figure}
    \centering
    \includegraphics[width=0.7\textwidth]{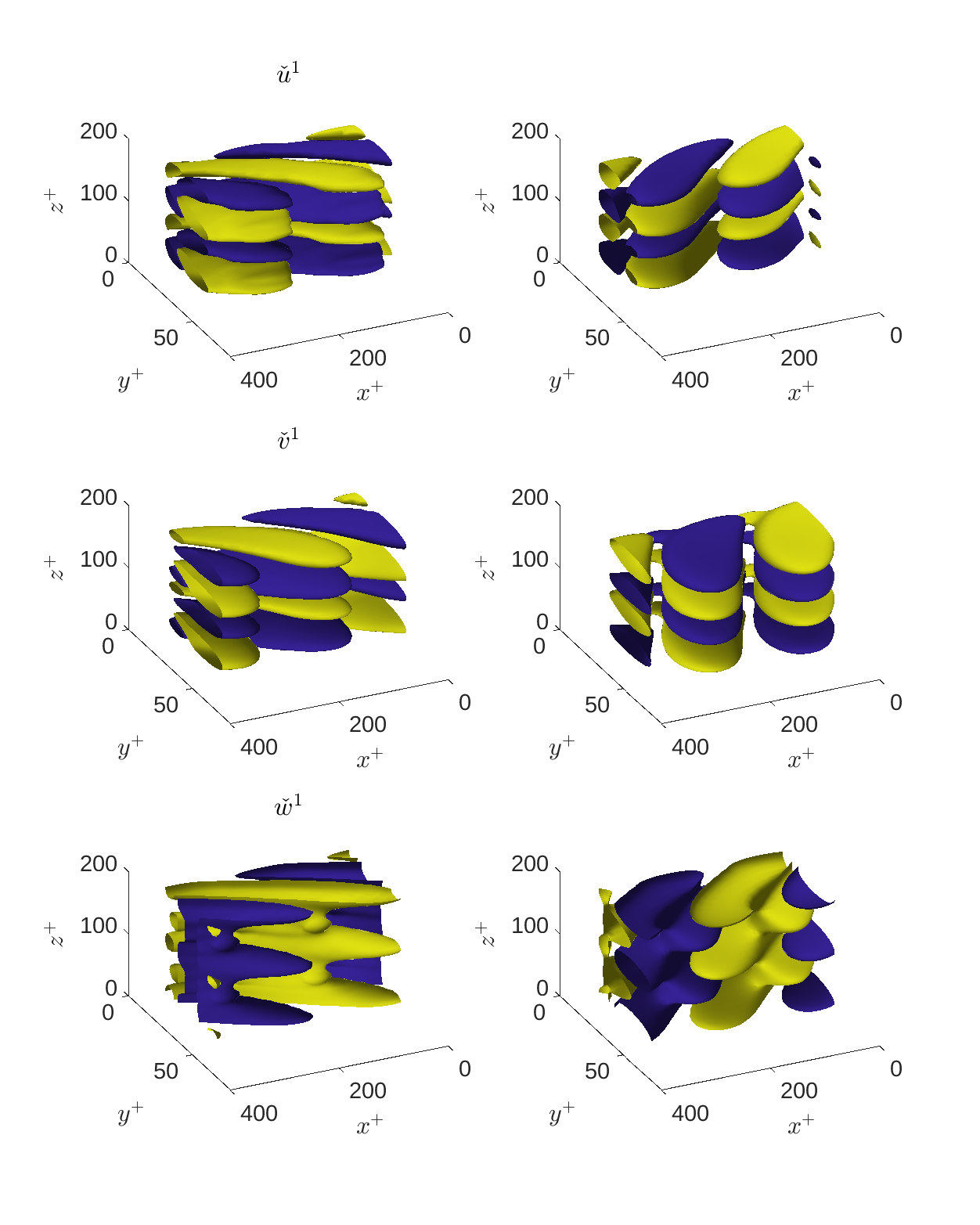}
    \caption{Optimal secondary transient growth modes with $\lambda_x^+=337$ at $Re_{\tau}=180$, primary streak amplitude $A=\bar{A}$ and $t_e$. Shown for minimal channel streaks with $\lambda^+_z=168$.
    Optimal input velocities are shown on the left column and optimal output velocities are shown on the right column.    
    The blue and yellow isosurfaces show $\pm0.2$ of the maximum of each velocity component. The Orr mechanism is noticeable in wall-normal velocity component where the structures become perpendicular to the wall. 
    Note, the wall is located vertically on the page at $y^+=0$ with the other at $y^+=2Re_\tau=360$, and the mean flow direction is from right to left. }
    \label{fig:opt_modes}
\end{figure}

Figure \ref{fig:opt_modes} illustrates a typical secondary optimal example at streamwise wavenumber $m=1$ ($\lambda_x^+=337$). The initial state of the optimal is shown in the left column and the final evolved state in  the right column. Similar to the optimal modes found around the laminar velocity profile with a spanwise streak \citep{cossu_et_al_2007}, the optimal modes are tilted upstream at the start of the process (flow is from right to left) and are then tilted downstream at the time of the maximum energy gain, with the wall-normal velocity component almost perpendicular to the wall. This is indicative  of the Orr mechanism.

%
%
%
%
 \begin{figure}
    \centering
    \includegraphics[width=0.32\textwidth]{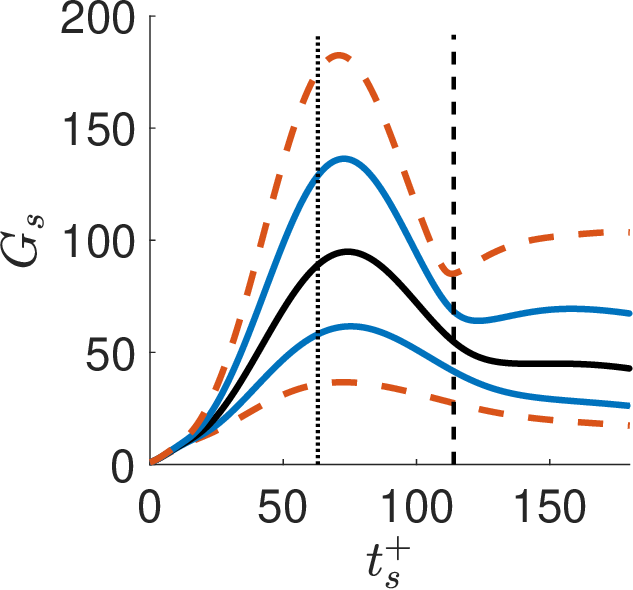}
     \includegraphics[width=0.32\textwidth]{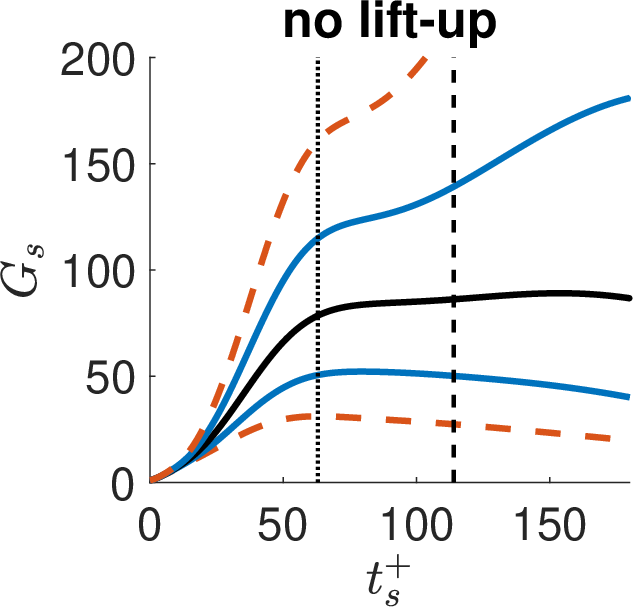}
      \includegraphics[width=0.32\textwidth]{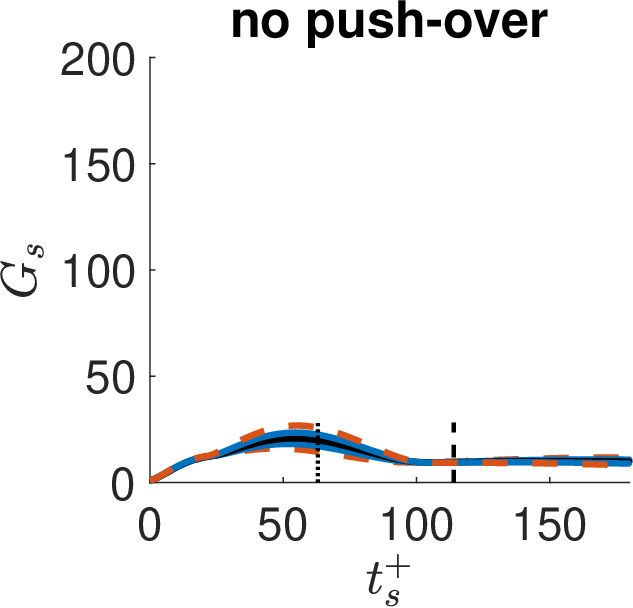}
    \caption{Secondary transient growth of perturbations with $\lambda_x=L_x$ ($m=1$) at $Re_{\tau} = 180$ for $\lambda_z^+=168$. Energy gain $G_s(t_s;A,180)$ is shown for different primary streak amplitudes $A=[\bar{A},\bar{A}\pm \sigma,\bar{A}\pm 2\sigma]$ (solid black, solid dark blue and dashed orange lines) 
    and found a) with full linear mechanism b) excluding `lift-up' c) excluding `push-over'. Dotted vertical lines show $t^+_{LD21} = 63$ time horizon while dashed vertical lines show the eddy turnover time $t^+_s = t^+_e$. }
    \label{fig:pushover}
\end{figure}

A key finding from LD21 (see their \S6.4) is that the `lift-up' mechanism is not important for the growth they observe but the `push-over' (see equation (6.24) in LD21) and Orr mechanisms are. To test this in our calculations, we consider the growth for the specific case of $m=1$ ($\lambda_x^+=337$) which just fits in the minimal channel and so was considered by LD21, and suppress either the lift-up or push-over mechanisms in turn: see Fig. \ref{fig:pushover}. With both present, the growth can range from 30 to 175 as the streak amplitude increases from $\bar{A}-2\sigma$ to $\bar{A}+2\sigma$ and the maximum growth at $A=\bar{A}$ is observed just beyond  $t^+_{LD21}=63$. 
These results are in good agreement with LD21 where energy gains of the frozen-in-time-streak field reach $\mathcal{O}(100)$  (see discussion surrounding their Figure 9). Further agreement with LD21 is found when considering the effect of lift-up and push-over mechanisms on the optimal gain values. While excluding the lift-up mechanism maintains the same sort of secondary transient growth factors, excluding the push-over mechanism causes the maximum optimal gain to collapse to $G_s\lesssim 25$ independent of the streak amplitude and decreasing the optimal timescale to just $t^+_s=36$ (the same effect was also seen in the larger channel to be discussed in \S \ref{3_3} ). We thus also find that the `push-over' mechanism is essential to obtain sufficient levels of transient growth needed to sustain turbulence whereas the lift-up mechanism is not.

The conclusion of this section is then that  a simplistic two-stage optimisation procedure is able to capture both the levels of growth found by LD21 in their DNS and the importance/unimportance of the push-over/lift-up mechanism in this. 
In fact, one could perhaps argue that {\em too} much growth is available through this simplistic approach. These are maximums, however, rather than expected values given generic, non-optimal initial secondary perturbations \citep[a point well made recently in ][]{Towne23}. The questions now are a) how do the possible growths vary with $\Rt$? and b) how do things change with a larger channel? In the latter, for example, it is unlikely that the most energetic streak will be that with the largest spacing which fits into a larger domain.

\subsection{Higher $\Rt$}              

\begin{figure}
\includegraphics[width=0.49\textwidth]{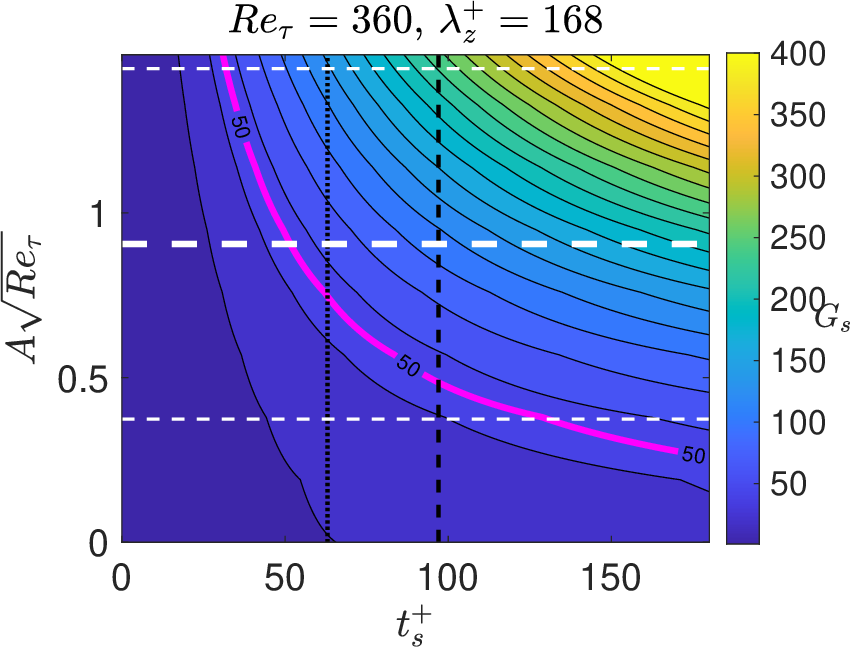}
\includegraphics[width=0.49\textwidth]{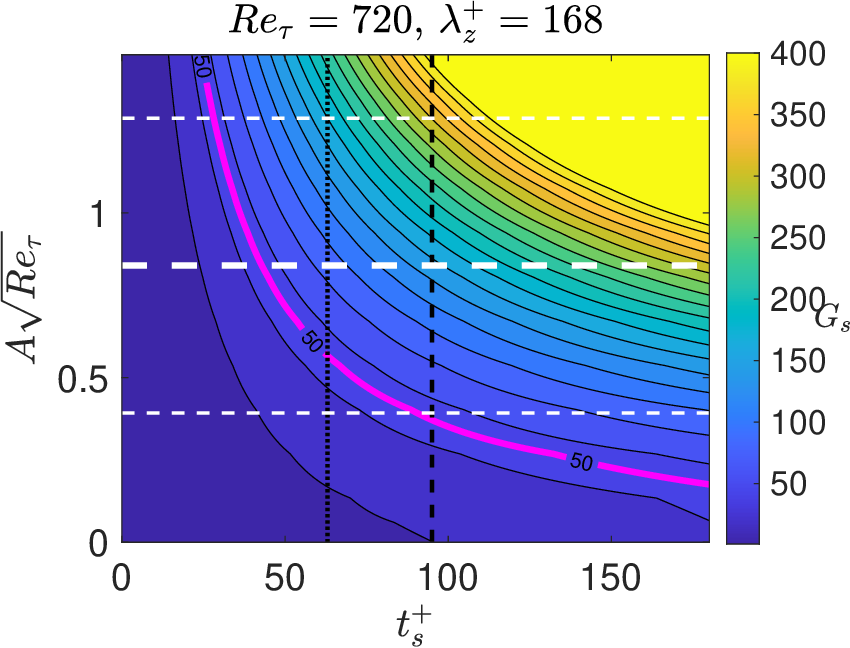}    
\caption{Secondary growth $G_s$ at $Re_{\tau}=360$ (left) and $Re_{\tau}=720$ (right) for the optimal streak $\lambda_z^{p+}=168$.   Time horizons $t_{LD21}^+$ and $t_e^+$ are marked with vertical black dotted and dashed lines respectively.     Horizontal white dashed lines show $\bar{A}\sqrt{Re_{\tau}}$ and $\left(\bar{A}\pm \sigma\right)\sqrt{Re_{\tau}}$.   Magenta solid contour shows the $G_s=50$ threshold and contour levels have $\Delta G_s =20$.}
    \label{fig:maxG}
\end{figure}

To explore higher $\Rt$ without changing the domain, the two-stage optimisation procedure was repeated at $\Rt=360$ and $\Rt=720$ in the minimal channel. 
The optimal primary streak remains $\lambda_z^{p+}=168$ and the corresponding contour plots of the optimal transient growth on the $(t_s^+,A\sqrt{Re_{\tau}})$ plane are shown in Figure \ref{fig:maxG} for relevant primary streak amplitudes observed in the simulations. Qualitatively, the plot remains largely unchanged when $\Rt$ is increased: the observed mean streak amplitude and $t_e^+$ both decrease but the secondary growth remains similar. In particular

\begin{equation}
G_s(t_{LD21}^+;168,\bar{A},720) \approx 90
\end{equation}
where
$m_{max}=1$ over integer values of $m$.
Again  the $G_s = 50$ contour threads the region around $(t_{LD21}^+,\bar{A}\sqrt{Re_\tau})$ with the main change being that  the contours become steeper for increasing $Re_\tau$.

%
%

%
\subsection{\label{3_3} Larger Geometry}        

Here, we use the  DNS data from the larger channel studied in LM15 (see Table \ref{tb:LM15} for details) to investigate the presence of larger lengthscales in the two-stage optimisation procedure (e.g. the larger channel is over an order of magnitude wider). The mean profiles shown in Figure \ref{fig:profiles}(top) are fairly similar for $\Rt=180$ (black solid line versus magenta dotted line)  but, as expected, start to deviate going towards the centre of the channel for higher $\Rt$: compare the profile at $\Rt=550$ for the larger channel (green dotted line) with the minimal channel profile at $\Rt=720$ (blue solid line). 
Relative to the minimal channel, the eddy turnover time function $t_e^+(y^+)$ intersects with the optimal streak function at a streak position slightly further from the wall and at a reduced turnover time $t_e^+(y^+_*)$ of 79 for $\Rt=180$ and 89 for $\Rt=550$: see figure \ref{fig:eddy}. As a consequence, the resulting optimal primary streaks while similar in structure  (upper plot in figure \ref{fig:profiles}) differ in spanwise wavelength: $\lambda_z^{p+}=106$ and 112 in the larger channel (so closer to the observed peak spacing of 100 \citep{Kline67}) for $\Rt=180$ and 550 respectively as opposed to $168$ in the minimal channel.

%
%
 \begin{figure}
    \centering  
\includegraphics[width=0.48\textwidth]{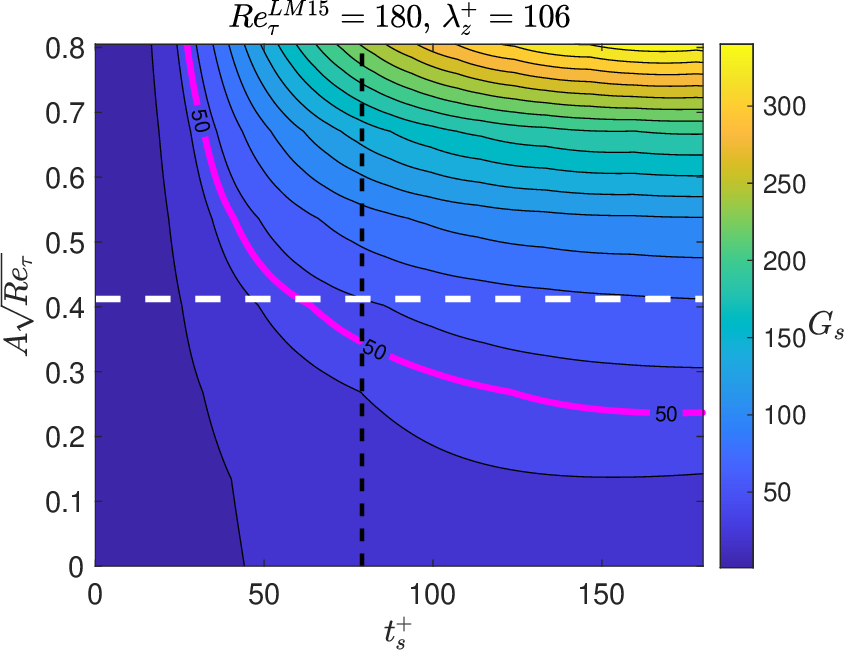}
\includegraphics[width=0.48\textwidth]{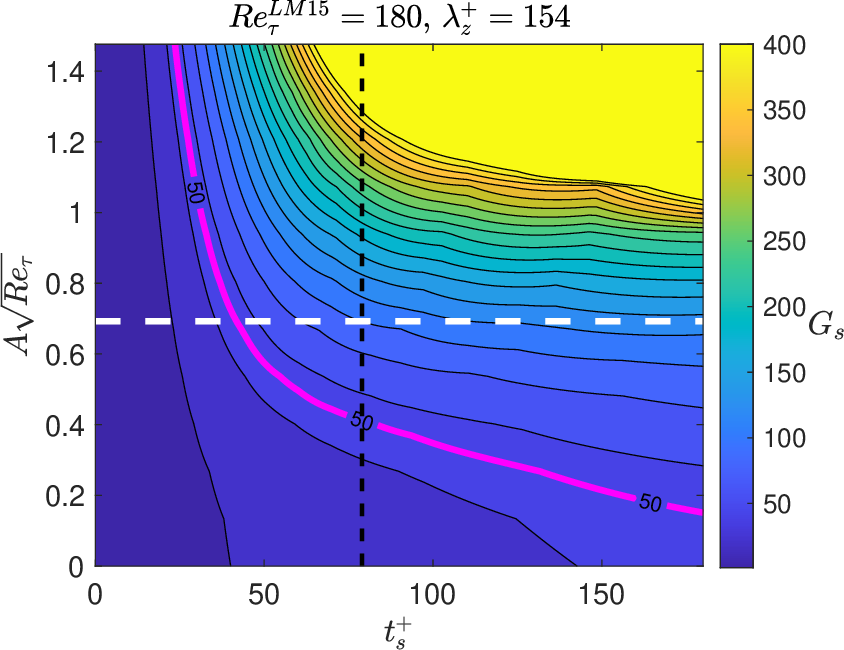}
\includegraphics[width=0.40\textwidth]{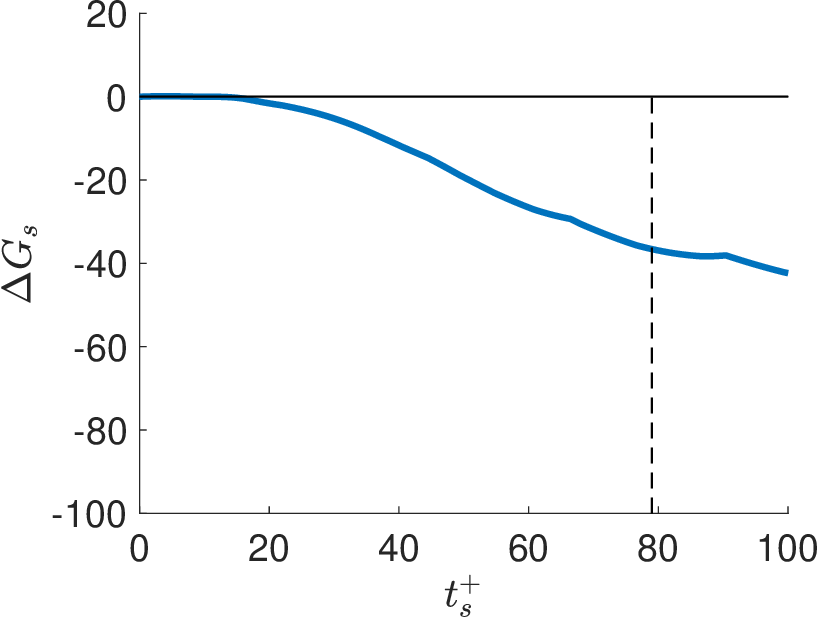}
    \caption{Top left (right):  Secondary transient growth $G_s$ contour plot over the $(t_s^+,A\sqrt{Re_{\tau}})$ plane at $Re_{\tau}=180$ for the LM15 channel and $\lambda^{p+}_z=106$ ($\lambda^{p+}_z=154$).  The optimisation is done for $m \in \{0,1,\ldots,25\}$ which is chosen so the smallest streamwise wavelength considered is approximately that used in the minimal channel.  
    Time horizon $t_e^+$ is marked with vertical black dashed lines, the magenta solid contour shows $G_s=50$ and the contour level spacing is $\Delta G_s =20$.
    Bottom: secondary transient growth difference 
    $\Delta G_s = G_s(t_e^+,106, \bar{A}_{106},180) - G_s(t_e^+,154,\bar{A}_{154},180)$ where $\bar{A}_{\lambda_z}$ is the observed mean amplitude of streak with spanwise wavelength $\lambda_z$. This shows that the $\lambda_z^{p+}=154$ streaks have larger secondary transient growth than the optimal streaks with $\lambda_z^{p+}=106$. }
    \label{fig:large_growth}
\end{figure}

The large channel, however, does show an interesting new feature compared to the minimal channel: the primary streak which produces the most secondary growth over all streaks of the {\em same} amplitude is not the streak of largest amplitude seen in the DNS (the streak energy was computed by integrating up to $y^+ \approx 18$ to focus on the near-wall structures). In fact adjusting for this difference in amplitude, the most energetic observed streak produces the most growth on account of its enhanced amplitude. At $Re_\tau=180$, the optimal (at fixed amplitude) primary streak has $\lambda_z^+=106$ whereas the largest observed amplitude streak has $\lambda_z^+=154$. This latter most energetic streak spacing  is essentially the same as that found in the minimal channel once the exact geometries are taken into account (in this larger channel, $\lambda_z^+=154$ has 11 wavelengths in the domain as opposed to $\lambda_z^+=169.6$ which has 10). Figure \ref{fig:large_growth} shows the optimal secondary growth over  the $(t_s^+,A \sqrt{\Rt})$ plane for both. The contour plots look very similar to the minimal channel results in figure \ref{fig:sg_180_minimal} with comparable growth possible, e.g.
\begin{equation}
G_s(t_e^+;106,\bar{A}_{106},180) \approx 60 \quad \& \quad G_s(t_e^+;154,\bar{A}_{154},180) \approx 100
\end{equation}
Here a subscript on $\bar{A}$ is now used to identify the streak wavelength and the eddy turnover time assumed for comparison purposes instead of $t^+_{LD21}$ which is only relevant for the minimal box. At $\Rt=180$, working in a larger domain does not change the conclusions drawn at the end of \S\ref{S3.1}.

\begin{figure}
    \centering  
\includegraphics[width=0.6\textwidth]{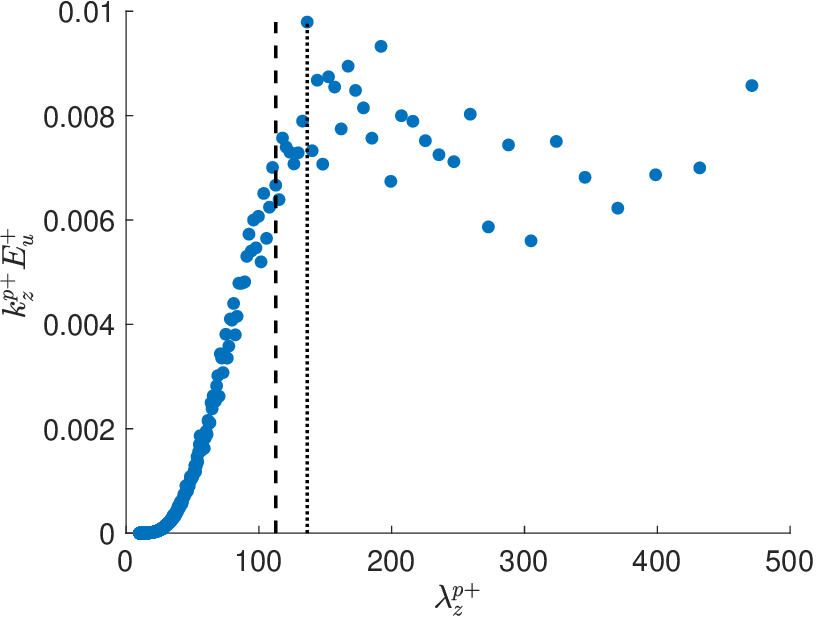}
    \caption{Streak energy $E_u^+$ premultiplied by $k_z^{p+}=2\pi / \lambda_z^{p+}$ against streak wavelength $\lambda_z^{p+}$ in LM15 simulations shown for $Re_{\tau}= 550$. The streak energy was integrated up to $y^+ \approx 18$ to focus on the near-wall structures only. Vertical dashed line shows the optimal streak with $\lambda_z^{p+}=112$ and vertical dotted line shows the most energetic streak with $\lambda_z^{p+}=136$.}
    \label{fig:LM15_streak_energy}
\end{figure}

\begin{figure}
    \centering    
\includegraphics[width=0.49\textwidth]{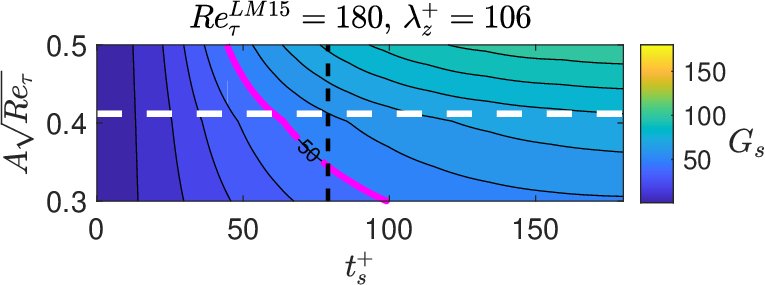}
\includegraphics[width=0.49\textwidth]{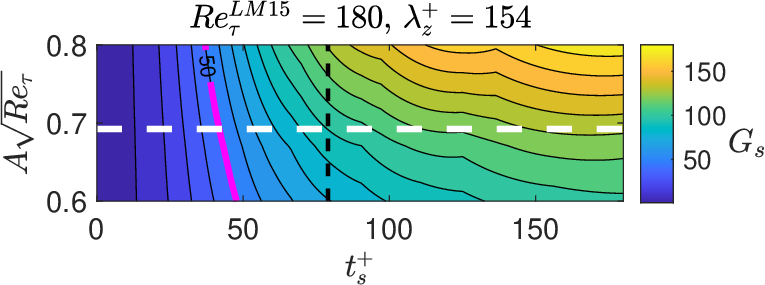}
\includegraphics[width=0.49\textwidth]{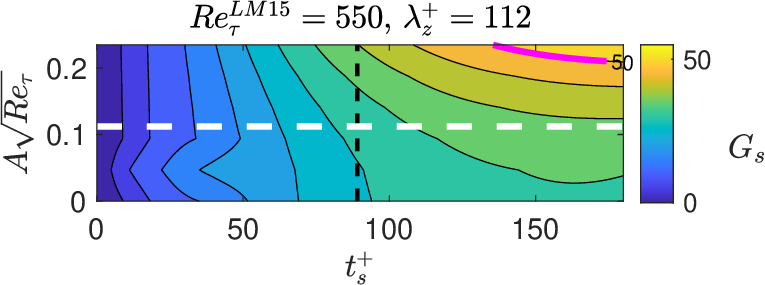}
\includegraphics[width=0.49\textwidth]{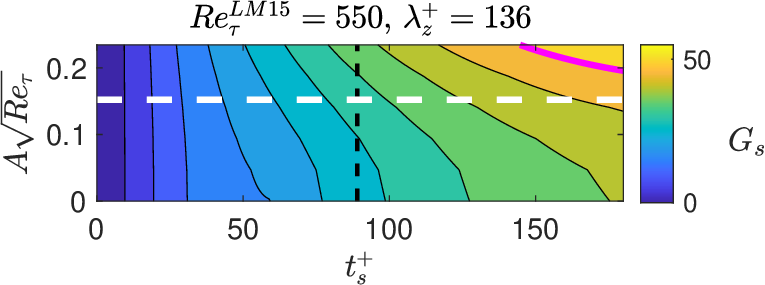}
    \caption{Secondary transient growth for the large LM15 channel. Contour levels every $\Delta G = 10$ for the top $Re_{\tau} =180$ plots and every $\Delta G = 5$ for the  $Re_{\tau} =550$ plots. The top plots are zoom ins of  the results in figures \ref{fig:large_growth}  to aid comparison. The magenta curve is the $G_s=50$ contour.}
    \label{fig:LM15}
\end{figure}

Moving to a higher $\Rt$, the primary optimal streak has $\lambda_z^{p+}=112$ in the large channel at $\Rt=550$ whereas the streak of largest mean amplitude has $\lambda_z^{p+}=136$: see figure \ref{fig:LM15_streak_energy}. Zoomed-in plots of the secondary growth possible over time and streak amplitude for these are shown in figure \ref{fig:LM15} along with the those for $\Rt=180$ for comparion. The key secondary growth estimators are 
\begin{equation}
G_s(t_e^+;112,\bar{A}_{112},550) \approx 31
\end{equation}
and
\begin{equation}
G_s(t_e^+;136,\bar{A}_{136},550) \approx 33
\end{equation}
 which are noticeably both reduced from $\Rt=180$ and much more similar to each other. Assuming this two-stage optimisation {\em is} capturing the correct behaviour, this indicates that the required energy growth in the near-wall region needed to sustain turbulence is a decreasing function of $\Rt$.
 Beyond appealing to reduced streak amplitudes and shorter times available for growth as $\Rt$ increases, a convincing explanation for this decrease needs to await some understanding of how the Orr and push-over mechanisms work together.
 
 The fact that the maximum secondary growth for the two primary streaks is much closer
 at $\Rt=550$ than $\Rt=180$ can partially be explained by the convergence of their spanwise wavelengths. But it is more likely that the growth is becoming insensitive  to the exact spanwise spacing because more and more degrees of freedom becoming active as $\Rt$ increases. 

%
%

%
%

\section{Discussion}
\label{sec:4}

%
%

In this study we have attempted to dissect the recent observation made in \cite{Lozano-Duran21} that viable streamwise-averaged flows in minimal channels achieve a threshold level of rapid transient growth. By examining a relatively simple two-stage  model of primary transient streak growth on a spanwise-invariant mean and then secondary transient growth on the now presumed frozen streaky primary flow, energy gain levels, timescales and the importance of the `push-over' mechanism were all found to be consistent with the results of LD21 at $Re_\tau=180$ in their minimal channel. 
Repeating this analysis for a larger channel at $\Rt=180$ using the database of \cite{lee_moser_2015} (LM15) reveals a discrepancy between the predicted optimal primary streak and the observed streak of largest energy, with latter producing most secondary growth on account of its larger amplitude. This situation is repeated at $\Rt=550$ but now the optimal streak and the most energetic streak have comparable spanwise spacing and the secondary growth possible on each is very similar albeit decreased from the $\Rt=180$ predictions.

%
%
The main benefit of our modelling approach is its interpretability and  the ability to look at larger domains and higher $Re_\tau$ relatively cheaply as done here. A DNS needs to be run to estimate the mean profile $U(y)$ (although a parameterised profile would probably suffice), the eddy turnover function $t_e^+(y^+)$ and the streak amplitudes. After that, the rest of the calculations are based on the spectral properties of appropriate linear operators. In particular, onerous time-dependent optimisation calculations are avoided. There is no denying, however, that the two-stage optimisation procedure explored here is  the simplest model  which could be imagined relevant to LD21 and is admittedly heuristic. The primary streaks are imagined to grow over a time $t_e^+$ and then artificially `frozen' so a secondary transient growth calculation can be performed. This has the advantage of involving only two linear computations performed here using standard matrix algebra albeit with the necessity of inserting the observed amplitude for the streak halfway through the process. A more realistic calculation would be to do a full optimisation problem where the secondary growth is optimised over all possible streak structures again frozen in time. A further step closer to reality would be to allow the streaks to evolve and then even allowing the growing secondary perturbations to influence the streak evolution. These problems are, of course, much more intensive computations seeking to get  ever closer to the  underlying simulations but at the same time run the risk of losing interpretability as they complexify. Nevertheless, there is no doubt a more joined-up optimisation procedure would be worth pursuing. The recent work computing growth for finite amplitude disturbances \citep[e.g.][]{Kerswell_18} points the way forward for examining the growth possible from the mean flow $U(y)\bx$ in which the streak field emerges as part of the optimisation along with the 3D fluctuation field. It would be fascinating to see if  the optimal which emerges from this one-step optimisation resembles that seen in simulations. 

The work described here has focussed upon the near-wall processes studied by LD21, but there is work which suggests that outer large scale structures are independently sustaining \citep{HwangCossu10}. Considering this region would certainly be worthwhile to see if it provides a lower energy growth threshold for the mean profile to satisfy. Even if it did, however, such growth would surely only be accessed by much more energetic fluctuations to the mean profile compared to the near-wall region and hence be much harder to initiate.

Another interesting direction for development is trying to estimate {\em expected} rather than maximum growth values. LD21 test an ensemble of streamwise-averaged DNS velocity fields for energy growth capability and folding in more of the properties of this into the secondary growth problem would clearly be valuable. A recent study by \cite{Towne23} has started to consider how the distribution of fluctuations available to initiate growth can influence the expected growth later. Here, the more relevant calculation would be  to study what the expected secondary growth characteristics are on the observed primary streak distribution.

Mechanistically, \cite{Lozano-Duran21} speculate that both the Orr and `push-over' effects are needed for sustaining turbulence in minimal channels and this is certainly replicated in our secondary growth analysis. However, these two processes appear to act over very different timescales: while the Orr mechanism acts on the fast advective timescale ( $\mathcal{O}(0.15 h/ u_{\tau})$ found by \cite{Jimenez15} in channel flow), `push-over' would seem  to act on the slow viscous timescale in analogy with `lift-up'. That is, slowly decaying streamwise rolls {\em spanwise}-advecting the now {\em spanwise} shear  over time to create streaks (for `lift-up, replace `spanwise' with 'cross-stream'). It is unclear  how Orr and push-over  synergise in a linear calculation to create optimal growth that cannot be achieved by only one of the two (the ability for one to feed energy into the other is clear from nonlinear calculations, e.g. see the appendix B in \cite{Kerswell_14}). The presence of shear in two different directions is surely significant and a study in the spirit of \cite{Jiao21}, who recently looked at the synergy between the `lift-up' and Orr processes, could clarify this.

%
%
Finally, we remark that the observation of  \cite{Lozano-Duran21} and our work here suggests a very plausible correction to  Malkus's (1956) hypothesis that turbulent mean profiles are marginally stable. Here, Malkus's core idea was  there should be many degrees of freedom (or fluctuations) just sustained but none growing on long times scales. Instead, as discussed above, the more relevant property of a realised mean flow profile could be that it can sustain {\em enough} transient energy growth of fluctuations over {\em short} (inertial) time scales. Importantly, this has to be  a {\em nonlinear} transient growth property of the mean profile since a finite-amplitude streak field (the primary phase above) has to be generated from the initial condition which can then itself help feed energy into a co-developing  3D fluctuation field (the secondary phase above).  Understanding the relevant initial amplitude criterion for this nonlinear growth  is an interesting issue as is examining how well our  deconstructed 2-stage process approximates the full nonlinear transient growth optimisation. We hope to report on these in the near future.

\vspace{1cm}
\backsection[Acknowledgement]{The authors are very grateful to Myoungkyu Lee for generously sharing data from his large channel simulations (Lee \& Moser 2015) and the referees for their interest in carefully examining the manuscript.}


\backsection[Funding]{VKM acknowledges financial support from EPSRC through a PhD studentship.}
\backsection[Declaration of interests]{ The authors report no conflict of interest.}

\bibliographystyle{jfm}
\bibliography{references}
\end{document}